\documentclass[aip,cha,numerical,reprint,eqsecnum,floats]{revtex4-1} 

\usepackage[english]{babel}
\usepackage{amsmath}
\usepackage{amssymb}
\usepackage{mathrsfs}
\usepackage{amsthm}
\usepackage{textcomp}
\usepackage{xcolor}
\usepackage{graphicx}
\usepackage{enumerate}
\usepackage{mathtools}
\usepackage{hyperref}
\usepackage[title]{appendix}
\usepackage{bm}

\newcommand{\J}{{\cal J}}
\newcommand{\dt}{\frac{{\rm d}}{{\rm d}t}}

\begin{document}
\title{Dynamical equivalence between Kuramoto models with first- and higher-order coupling}
\author{Robin Delabays}
\affiliation{School of Engineering, University of Applied Sciences of Western Switzerland, CH-1950 Sion, Switzerland}
\affiliation{Institut f\"ur Automatik, ETH Z\"urich, CH-8092 Z\"urich, Switzerland} 

\date{\today}

\begin{abstract}
 The Kuramoto model with high-order coupling has recently attracted some attention in the field of coupled oscillators in order, for instance, to describe clustering phenomena in sets of coupled agents. 
 Instead of considering interactions given directly by the sine of oscillators' angle differences, the interaction is given by the sum of sines of integer multiples of the these angle differences.  
 This can be interpreted as a Fourier decomposition of a general $2\pi$-periodic interaction function. 
 We show that in the case where only one multiple of the angle differences is considered, which we refer to as the \emph{Kuramoto model with simple $q^{\rm th}$-order coupling}, 
 the system is dynamically equivalent to the original Kuramoto model. 
 In other words, any property of the Kuramoto model with simple higher-order coupling can be recovered from the standard Kuramoto model. 
\end{abstract}

\maketitle

\begin{quotation}
 Along the last decades, the Kuramoto model has attracted a lot of interest in the field of synchronization of coupled oscillators. 
 Its simple formulation and the variety of synchronization phenomena that it describes make it a good candidate to investigate such phenomena both numerically and analytically. 
 The sinusoidal interaction in the Kuramoto model can be seen as the first order of the Fourier decomposition of a general coupling function. 
 It is then natural to extend the coupling to higher orders, such that the dynamics depend on the sines of multiples of the angle differences. 
 In this manuscript, we show that considering only the angle difference or a unique integer multiple of it in the sinusoidal coupling is qualitatively equivalent. 
\end{quotation}

\section{Introduction}
In the context of synchronization of coupled dynamical systems, the Kuramoto model~\cite{Kur75,Kur84} has drawn a lot of attention within the last decades.~\cite{Str00,Ace05,Dor14} 
Synchony is observed in many real systems, ranging from the brain's oscillatory pacemaker cells establishing the circadian rythm,~\cite{Lu16} 
to synchronous machines connected to the high-voltage AC electrical grid.~\cite{Ber00,Dor13} 
This popularity brought the topic to a point where the remaining open questions are both hard and poorly rewarding to answer. 
To describe more realistic systems, some generalized versions of the Kuramoto model have been considered, as, for instance, meshed interaction graphs, higher order dynamics,~\cite{Dor13} 
or directed interactions.~\cite{Res06,Del19}
One of these generalizations is to consider higher-order couplings.~\cite{Han93,Dai96,Ska11,Kom13,Li14,Wan17,Eyd17,Li19a}
While the Kuramoto model is defined as 
\begin{align}\label{eq:kuramoto1}
 \dot{\theta}_i &= \omega_i - \frac{K}{n}\sum_{j=1}^n\sin(\theta_i-\theta_j)\, ,  
\end{align}
for $i\in\{1,...,n\}$, where $\theta_i\in\mathbb{R}$ is the $i^{\rm th}$ oscillator's angle, $\omega_i\in\mathbb{R}$ is its natural frequency, and $K>0$ is the coupling strength, 
the \emph{Kuramoto model with $q^{\rm th}$-order coupling}, for $q\in\mathbb{Z}_{>0}$, is defined as 
\begin{align}\label{eq:kuramotoQ}
 \dot{\theta}_i &= \omega_i - \sum_{j=1}^n\sum_{\ell=1}^q \frac{K_\ell}{n}\sin\left[\ell\cdot(\theta_i-\theta_j)\right]\, ,
\end{align}
for $i \in \{1,...,n\}$. 
The sum over $\ell$ can be seen as a truncated Fourier decomposition of a general $2\pi$-periodic coupling function. 

To this day, most of the works about this version of the model has been limited to second-order couplings ($K_{1,2}\neq0$), 
which already exhibits behaviors significantly different from the original Kuramoto model.~\cite{Kom13,Li14,Wan17,Eyd17} 
While most descriptions of the synchronous states of this model, for $q=2$ and large $n$, have been performed numerically,~\cite{Li14,Wan17} 
an analytical approach, based on a self-consistency equation for the order parameters of the system, is given in Ref.~\onlinecite{Kom13}. 
The Kuramoto model with $q^{\rm th}$-order coupling has been used to describe clustering phenomena in systems of coupled synchronized oscillators. 
Clustered synchronous states of some particular versions of the Kuramoto model with second-order coupling and their dynamics are described in Refs.~\onlinecite{Han93,Dai96}. 

When one of the coupling orders largely dominates the others, a simplifying assumption is to consider the case where $K_\ell=0$ for all $\ell\neq q$, 
which we refer to as the \emph{Kuramoto model with simple $q^{\rm th}$-order coupling}.~\cite{Ska11,Li19a} 
In this case, Eq.~\eqref{eq:kuramotoQ} reduces to  
\begin{align}\label{eq:kuramotoq}
 \dot{\theta}_i &= \omega_i - \frac{K}{n}\sum_{j=1}^n\sin\left[q\cdot(\theta_i-\theta_j)\right]\, ,
\end{align}
for $i \in \{1,...,n\}$, which is the dynamical system considered in this manuscript. 
Note that the Kuramoto model, Eq.~\eqref{eq:kuramoto1}, can be seen as the Kuramoto model with simple first-order coupling. 
An analytical description of the transient dynamics of Eq.~\eqref{eq:kuramotoq} and its synchronous clustered states are given in Ref.~\onlinecite{Ska11}. 

In this manuscript, we show that the dynamical systems described by Eqs.~\eqref{eq:kuramoto1} and \eqref{eq:kuramotoq} are qualitatively equivalent, in the sense that both have the same dynamical properties (fixed points, linear stability, basin stability, order parameter) up to a projection from the state space to itself. 
Doing so, we make rigorous the claim below Eq.~(3) in Ref.~\onlinecite{Li14} that \emph{``In the case with $K_1 = 0$ (or $K_2 = 0$), the model is reduced to the original [Kuramoto model]...''}. 
It also explains the striking similarity between Figure 2 in Ref.~\onlinecite{Li19a} (Kuramoto model with simple second-order coupling) and Figure 2 in Ref.~\onlinecite{Li19b} (Kuramoto model with simple first-order coupling). 
In Sec.~\ref{sec:equiv}, we see that a direct relation can be drawn between the two models, allowing to translate any property of one model to the other. 
Therefore, a thorough investigation of the Kuramoto model with simple $q^{\rm th}$-order coupling is not needed, as any of its properties can be derived from properties of the original Kuramoto model. 
A selection of such properties and some implication for potential applications are detailed in Sec.~\ref{sec:prop}.

{\bf Remark.}
\textit{In Eq.~(\ref{eq:kuramotoq}), we consider all-to-all coupling, and we will do so for the whole manuscript for sake of readability. 
Nevertheless, our results can be straightforwardly extended to any coupling graph. 
We comment on this at the end of Sec.~\ref{sec:prop}.}

\section{Dynamical equivalence}\label{sec:equiv}
The Kuramoto model is usually considered as a dynamical system on the torus $\mathbb{T}^n$. 
We consider the variables $\theta_i$ as elements of $\mathbb{S}^1$, which we parametrize as $[0,2\pi)$ with periodic boundary conditions. 
We show now that the two following dynamical systems 
\begin{align}
 \dot{\theta}_i &= q\omega_i - \frac{qK}{n}\sum_{j=1}^n\sin\left(\theta_i-\theta_j\right)\, ,\label{eq:k1} \\
 \dot{\theta}_i &= \omega_i - \frac{K}{n}\sum_{j=1}^n\sin\left[q\cdot(\theta_i-\theta_j)\right]\, . \label{eq:kq}
\end{align}
describe the same dynamics, up to rescaling the variables $\theta_i$ by a factor $q$. 
The rescaling is performed by the covering map from $\mathbb{T}^n$ to itself, which we define parametrically as 
\begin{align}
\begin{split}
 \pi_q \colon \mathbb{T}^n & \longrightarrow \mathbb{T}^n \\
 (\theta_1,...,\theta_n) & \longmapsto (q\theta_1,...,q\theta_n) \mod 2\pi\, ,
\end{split}
\end{align}
where the modulo is applied elementwise.

More formally, we show below that $\pi_q$ sends any solution of Eq.~\eqref{eq:kq} to a solution of Eq.~\eqref{eq:k1}, and that any lift of a solution of Eq.~\eqref{eq:k1} through $\pi_q$ is a solution of Eq.~\eqref{eq:kq}. 
As $\pi_q$ is a smooth covering map, all properties of a solution of Eq.~\eqref{eq:kq} are preserved in its projection [and similarly for the lift of a solution of Eq.~\eqref{eq:k1}]. 
This is what we mean by \emph{dynamical equivalence}. 

\subsection{Projecting}
Let $\Theta_{\bm{\psi}^*}\colon\mathbb{R}\to\mathbb{T}^n$ be the solution of Eq.~\eqref{eq:kq} with initial conditions $\bm{\psi}^*$. 
We verify that the projection $\pi_q\Theta_{\bm{\psi}^*}(t)$ solves Eq.~\eqref{eq:k1}, 
\begin{align}
 \dt\pi_q\Theta_{\bm{\psi}^*} 
 &= q\omega_i - \frac{qK}{n}\sum_{j=1}^n\sin\left[q\left(\Theta_{\bm{\psi}^*,i}-\Theta_{\bm{\psi}^*,j}\right)\right] \\
 &= q\omega_i - \frac{qK}{n}\sum_{j=1}^n\sin\left(\pi_q\Theta_{\bm{\psi}^*,i}-\pi_q\Theta_{\bm{\psi}^*,j}\right)\, ,
\end{align}
with initial conditions $\pi_q\Theta_{\bm{\psi}^*}(0)=q\bm{\psi}^*$. 

\subsection{Lifting up}
The other way is a bit more intricate, because there are multiple preimages through $\pi_q^{-1}$ for each element of $\mathbb{T}^n$. 
Let $\Phi_{\bm{\eta}^*}\colon\mathbb{R}\to\mathbb{T}^n$ be the solution of Eq.~\eqref{eq:k1} with initial conditions $\bm{\eta}^*\in\mathbb{T}^n$. 
The preimage $\pi_q^{-1}\bm{\eta}^*$ is a set of $q^n$ points, one of them being $q^{-1}\bm{\eta}^*\in[0,2\pi/q)^n\subset\mathbb{T}^n$, whose $i^{\rm th}$ component is $\eta_i/q$. 
The other $q^n-1$ can be constructed as 
\begin{align}
 q^{-1}\bm{\eta}^* + \frac{2\pi}{q}\bm{\rho}\, ,
\end{align}
with $\bm{\rho}\in\{0,1,...,q-1\}^n$. 
Each one of these points can be chosen as a starting point for the lifting. 

Now, whichever element $\bm{\psi}^*\in\pi_q^{-1}\bm{\eta}^*$ we choose as starting point for the lifting, there is a unique smooth lifting $\pi_q^{-1}\Phi_{\bm{\eta}^*}(t)$ of the trajectory $\Phi_{\bm{\eta}^*}(t)$ satisfying the two following properties: 
\begin{enumerate}[(i)]
 \item $\pi_q^{-1}\Phi_{\bm{\eta}^*}(0) = \bm{\psi}^*$;
 \item $\pi_q\left[\pi_q^{-1}\Phi_{\bm{\eta}^*}(t)\right] = \Phi_{\bm{\eta}^*}(t)$, for all $t\in\mathbb{R}$.
\end{enumerate}

The time derivative of the $i^{\rm th}$ component of this lifting is 
\begin{align}
 \dt & \left(\pi_q^{-1}\Phi_{\bm{\eta}^*}\right)_i = q^{-1}\dot{\Phi}_{\bm{\eta}^*,i} \\
 &= q^{-1}\left[q\omega_i - \frac{qK}{n}\sum_{j=1}^n\sin\left(\Phi_{\bm{\eta}^*,i}-\Phi_{\bm{\eta}^*,j}\right)\right] \\
 &= \omega_i - \frac{K}{n}\sum_{j=1}^n\sin\left[q\left(\pi_q^{-1}\Phi_{\bm{\eta}^*,i}-\pi_q^{-1}\Phi_{\bm{\eta}^*,j}\right)\right]\, .
\end{align}
The lifting then solves Eq.~\eqref{eq:kq} with initial conditions $\bm{\psi}^*$, and this is true independently of the choice of representative $\bm{\psi}^*\in\pi_q^{-1}\bm{\eta}^*$. 
The preimage, by $\pi_q^{-1}$, of a solution $\Phi_{\bm{\eta}^*}$ of Eq.~\eqref{eq:k1} is then a set of solutions of Eq.~\eqref{eq:kq} differing from one another by a shift $2\pi\bm{\rho}/q$, 
with $\bm{\rho}\in\{0,1,...,q-1\}^n$. 

\begin{figure*}
 \centering
 \includegraphics[width=.9\textwidth]{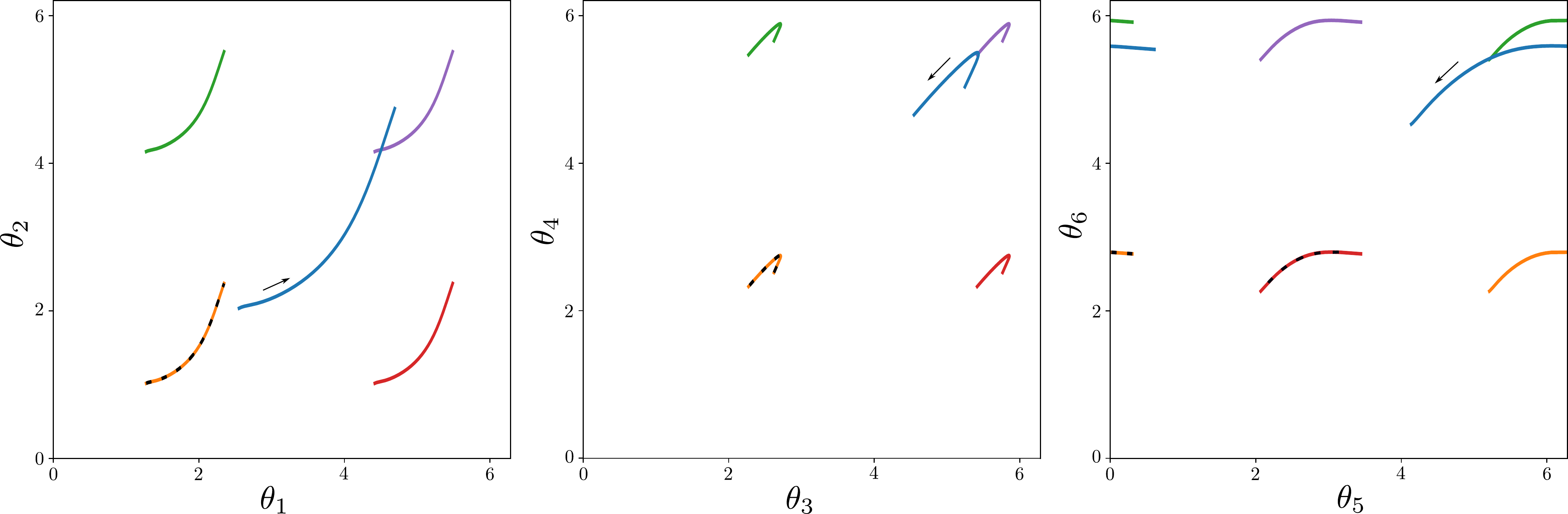}
 \caption{Example of the trajectories of the dynamical systems Eq.~\eqref{eq:k1} (blue) and Eq.~\eqref{eq:kq} (orange, green, red, and purple), with $n=6$ and $q=2$. 
 Boundary conditions are periodic. 
 Initial conditions were, respectively, $\bm{\psi}^*\in[0,2\pi/q)^n$ (orange, taken randomly), $\bm{\psi}^*+(\pi,0,\pi,0,\pi,0)$ (red), $\bm{\psi}^*+(0,\pi,0,\pi,0,\pi)$ (green), $\bm{\psi}^*+(\pi,\pi,\pi,\pi,\pi,\pi)$ (purple), and $q\bm{\psi}^*$ (blue). 
 The black dashed line is obtained by dividing all angles of the solution of Eq.~\eqref{eq:k1} (blue) by $q$. 
 The arrow shows the direction of the time evolution. 
 The solutions of Eq.~\eqref{eq:kq} are simply translations of each other, and the solution of \eqref{eq:k1} is a homogeneous dilatation of factor $q$ of the others. }
 \label{fig:equiv}
\end{figure*}

\subsection{Dynamical equivalence}
It is now clear that 
\begin{align}\label{eq:proj}
 \pi_q\left(\pi_q^{-1}\Phi_{\bm{\eta}^*}\right) &= \Phi_{\bm{\eta}^*}\, ,
\end{align}
and the unique smooth lifting of $\pi_q\Theta_{\bm{\psi}^*}$ such that $\pi_q^{-1}\left(\pi_q\Theta_{\bm{\psi}^*}\right)(0)=\bm{\psi}^*$ is exactly 
\begin{align}\label{eq:lift}
 \pi_q^{-1}\left(\pi_q\Theta_{\bm{\psi}^*}\right) &= \Theta_{\bm{\psi}^*}\, .
\end{align}
Properties of the solutions are then preserved by the projection $\pi_q$ as well as by its correponding lifting. 
We verify this for some dynamical properties in Sec.~\ref{sec:prop}.

\section{Consequences on clustering}\label{sec:prop}
Now that we established the dynamical equivalence between the Kuramoto model with simple first- and higher-order coupling, we review some results known for the original Kuramoto model and translate them in the Kuramoto model with $q^{\rm th}$-order coupling, in order to derive some results about clustering in the latter. 

\subsection{Fixed points} 
A given fixed point of the original Kuramoto model corresponds to $q^n$ fixed points of the Kuramoto model with $q^{\rm th}$-order coupling. 
Each of these fixed points differ by a shift $2\pi\bm{\rho}/q$, with $\bm{\rho}\in\{0,1,...,q-1\}^n$. 
When the natural frequencies are small ($\omega_i\ll K/n$), the synchronous state $\bm{\theta}^*\in[0,2\pi)^n$ of Eq.~\eqref{eq:k1} is such that all angles are close to each other. 
For any $\bm{\rho}\in\{0,1,...,q-1\}^n$, the point $q^{-1}\bm{\theta}^*+2\pi\bm{\rho}/q\in[0,2\pi)^n$ is a synchronous state for Eq.~\eqref{eq:kq}. 
The integer vector $\bm{\rho}$ describes the \emph{clustering pattern} of the corresponding synchronous state. 
If $\rho_i=\rho_j$, oscillators $i$ and $j$ are close to each other (at least for rather small natural frequencies), and approximately $2\pi/q$ appart from an oscillator $k$ such that $\rho_k=\rho_i\pm1$. 
Then each oscillators with the same value in $\bm{\rho}$ form a cluster. 
The number of different values in $\bm{\rho}$ gives the total number of clusters in the synchronous state under consideration. 
From the point of view of the dynamics however, the clustering pattern has no effect. 
The $2\pi/q$ shifts introduced by the vector $\bm{\rho}$ leave Eq.~\eqref{eq:kq} unchanged. 
The only information that is lost between the $q^{\rm th}$- and first-order coupling Kuramoto models is the clustering pattern. 
But this is only a combinatorial problem that can be addressed independently of dynamical considerations. 
In Fig.~\ref{fig:equiv}, for instance, each color corresponds to a different clustering pattern. 
At the end of the trajectory, the orange line has five oscillators in one cluster ($\{1,2,3,4,6\}$), with angles in $[0,\pi)$ and oscillator $5$ forming a cluster by itself, with its angle in $[\pi,2\pi)$. 
Similarly, for the green line, oscillators $1$ and $3$ are in one cluster with angles in $[0,\pi)$ and the others are in the cluster with angles in $[\pi,2\pi)$.

\subsection{Linear stability}
One can verify that the linear stability of Eq.~\eqref{eq:kq} at a fixed point $\bm{\theta}^*$ is identical to the linear stability of Eq.~\eqref{eq:k1} at $\pi_q\bm{\theta}^*$. 
More precisely, the Jacobian matrices $\J_q(\bm{\theta}^*)$ of Eq.~\eqref{eq:kq} at $\bm{\theta}^*$, and $\J_1(\pi_q\bm{\theta}^*)$ of Eq.~\eqref{eq:k1} at $\pi_q\bm{\theta}^*$ are equal. 
Thus, for a given distribution of natural frequencies, all clustered states have the same linear stability properties. 

\subsection{Order parameter}
The order parameter $r_1$ is a quantity describing the level of coherence between oscillators' angles in the Kuramoto model. 
It has been a major ingredient to analyze synchronization in this model.~\cite{Kur75,Aey04,Mir05,Ver08} 
To take clustering into account, in the Kuramoto model with higher-order coupling, it has been generalized~\cite{Ska11} to the $q^{\rm th}$ order parameter 
\begin{align}
 r_q(\bm{\theta}) &\coloneqq \frac{1}{n}\left|\sum_{j=1}^ne^{iq\theta_j}\right|\, .
\end{align}
It directly translates from the Kuramoto model with simple $q^{\rm th}$-order coupling to its equivalent Kuramoto model with first-order coupling. 
Namely, the order parameter $r_q(\bm{\theta}^*)$ is equal to $r_1(\pi_q\bm{\theta}^*)$. 
A large $q^{\rm th}$ order parameter $r_q$ indicates that the current state of Eq.~\eqref{eq:kq} is clustered, but does not give any information about the clustering pattern, because it is blind to any angle shift of $2\pi/q$. 
The order parameter $r_q$ takes the same value $r=r_1(\bm{\psi}^*)$ for each element of the preimage $\pi_q^{-1}\bm{\psi}^*$. 

More generally, $r_p\approx 1$ for $1\leq p\leq q$ indicates that the system is clustered in $p$ equidistance clusters. 
Furthermore, if $r_p\approx 1$, it implies that $r_{kp}\approx 1$ for any $k\in\mathbb{N}$. 
Looking at all the order parameters with $1\leq p\leq q$ can give more information about the clustering pattern. 
For instance in the special case where $q=2$, if $r_2=1$, then one can verify that the number of oscillators in each cluster $n_1$ and $n_2$ respectively are given by
\begin{align}
 n_1 &= \frac{(1+r_1)n}{2}\, , & n_2 &= \frac{(1-r_1)n}{2}\, .
\end{align}
In more general cases, however, it is not possible, as far as we can tell, to determine the number of oscillators in each cluster only based on the order parameters.

\subsection{Synchronization} 
It is known~\cite{Dor11} for the Kuramoto model, Eq.~\eqref{eq:k1}, that if the coupling is sufficiently large to grant the existence of a synchronous state [$K>\max_{i,j}(\omega_i-\omega_j)$], then there exists a value $\gamma_{\rm max}\in(\pi/2,\pi]$ such that the system synchronizes if all initial angles are in an arc of length at most $\gamma_{\rm max}$. 
In the Kuramoto model with simple $q^{\rm th}$-order coupling, this translates as follows. 
First, for the same value of $K$, the system synchronizes to the single cluster fixed point if all initial angles are in an arc of length $\gamma_{\rm max}/q$. 
Second, if the initial conditions $\bm{\psi}^*$ are such that all angles of $\bm{\psi}^*-2\pi\bm{\rho}^*/q$ are contained in an arc of length $\gamma_{\rm max}/q$, then the system synchronizes to the state with clusters given by $\bm{\rho}^*\in\{0,1,...,q-1\}^n$. 

\subsection{Basins of attraction}
\begin{figure}
 \centering
 \includegraphics[width=.45\textwidth]{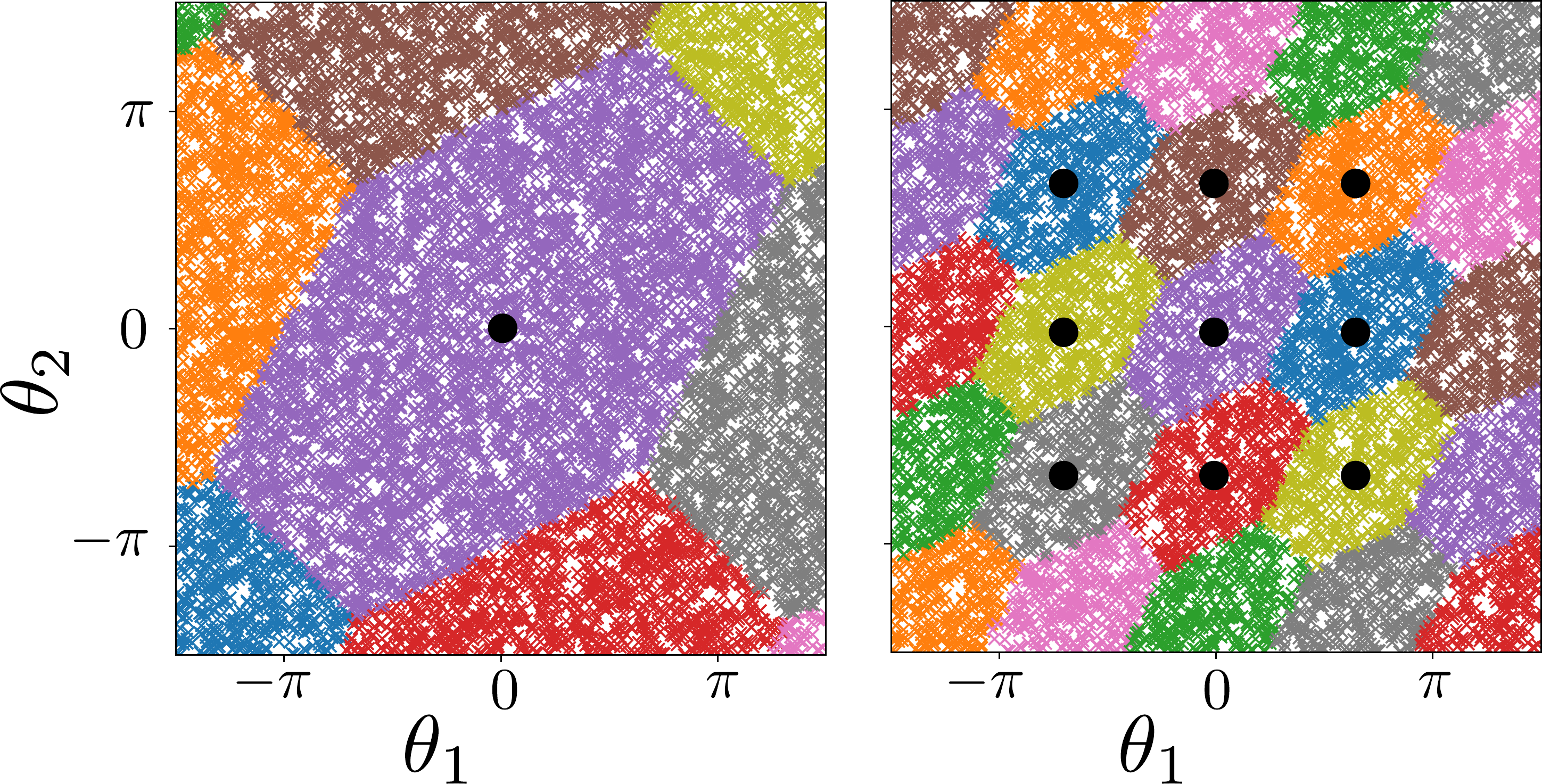}
 \caption{Basins of attraction of various synchronous states of Eq.~\eqref{eq:k1} (left panel) and Eq.~\eqref{eq:kq} (right panel) with $n=3$, $q=3$, and identical natural frequencies ($\omega_i\equiv0$). 
 The angle $\theta_3$ was fixed at $0$, to remove the degree of freedom corresponding to the constant angle shift. 
 Each figure is composed of 10'000 random initial conditions, and the color of each cross depends on the state to which it synchronizes. 
 For the Kuramoto model, Eq.~\eqref{eq:k1}, all initial condition converge to the same synchronous state (black dot, left panel), the colors only indicate if some angles accumulated a multiple of $2\pi$ due to the dynamics. 
 For the Kuramoto model with simple $q^{\rm th}$-order coupling, Eq.~\eqref{eq:kq}, there are nine different synchronous states (black dots, right panel), with different clustering pattern, the other basins correspond to translations of these nine basins. 
 This illustrates that the basins of Eq.~\eqref{eq:kq} are a copy of those of Eq.~\eqref{eq:k1} rescaled by a factor $1/3$.}
 \label{fig:basins}
\end{figure}

By equivalence of the dynamics, the basin of attraction of fixed point $\bm{\theta}^*$ of Eq.~\eqref{eq:kq} is a copy of the basin of attraction of the fixed point $\pi_q\bm{\theta}^*$ of Eq.~\eqref{eq:k1}, rescaled by a factor $q^{-1}$. 
Namely, its volume is $q^{-n}$ times the volume of the basin of attraction of $\pi_q\bm{\theta}^*$. 
We illustrate this in Fig.~\ref{fig:basins}, where we show the basins of attraction for the systems Eq.~\eqref{eq:k1} (left panel) and Eq.~\eqref{eq:kq} (right panel). 

\begin{figure*}
 \centering
 \includegraphics[width=.8\textwidth]{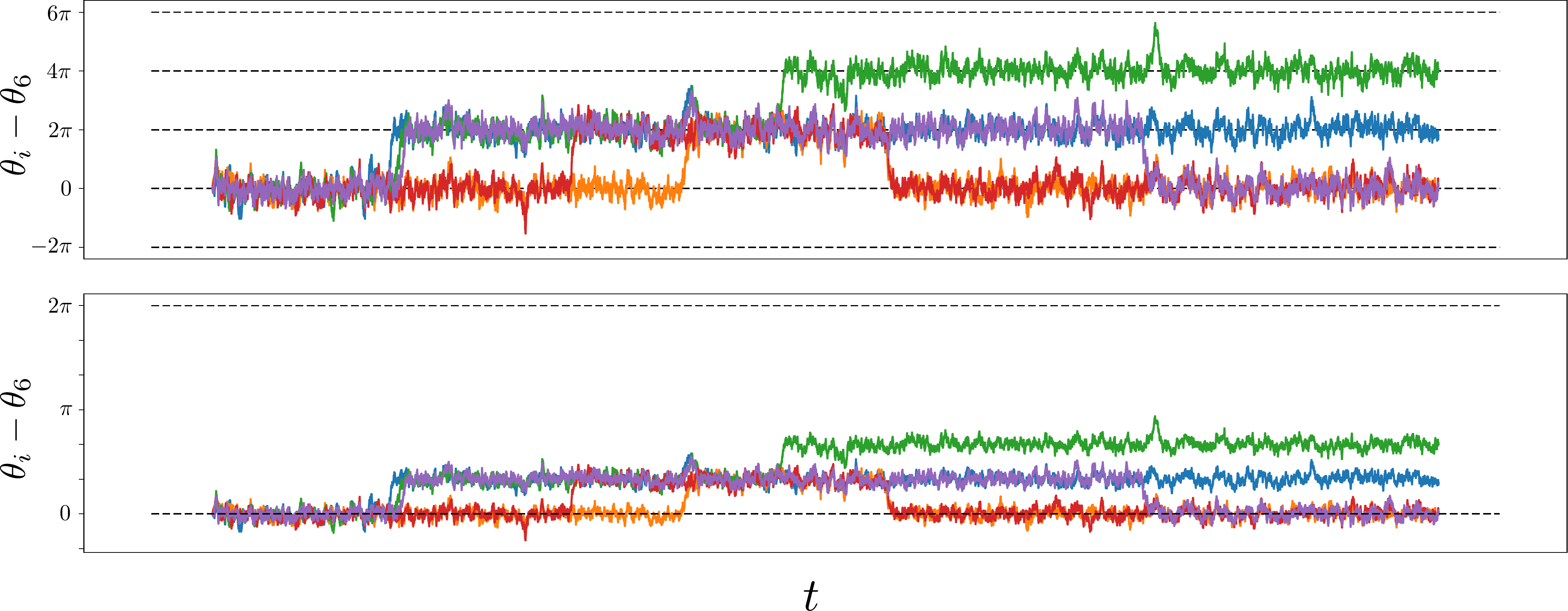}
 \caption{Time evolution of the angles in the Kuramoto model (top panel) and the Kuramoto model with simple $6^{\rm th}$-order coupling (bottom panel) with $n=6$, both subject to additive white noise. 
 Initial conditions are $(0,...,0)$ for both systems, natural frequencies are identically zero, and the coupling is $K_1=1$ and $K_6=1/6$ respectively. 
 The noise sequences are the same, with amplitude divided by $6$ for the Kuramoto model with $6^{\rm th}$-order coupling. 
 The black dashed lines indicate the mulitple of $2\pi$, i.e., the limits of the periodic boundary conditions. 
 After each jump in the Kuramoto model, the system converges back to its initial conditions, whereas with $6^{\rm th}$-order coupling, the Kuramoto model converges to a clustered state, even if the trajectories are qualitatively the same, this is only due to the different coupling functions.}
 \label{fig:jumps}
\end{figure*}

\subsection{Basin escape}
Suppose we introduce an additive noise in Eq.~\eqref{eq:kq} to account for unpredictable perturbation of the environment. 
This will eventually lead our system to jump from a synchronous state to another.~\cite{Dev12,Tyl19}
For the Kuramoto model, Eq.~\eqref{eq:k1}, such jumps bring the system from a synchronous state to a translate of itself, where some angles slip and accumulate integer multiples of $2\pi$. 
Lifting up such a trajectory to the Kuramoto model with simple $q^{\rm th}$-order coupling, the jumps then occur between the basins of attraction of different clustered states, where some angles accumulate an integer multiple of $2\pi/q$. 
Additive noise in Eq.~\eqref{eq:kq} is then a possible mechanism for cluster formation. 

If the noise has sufficiently small amplitude, the system remains for a long time in a neighborhood of a synchronous state, until the noise generates a sequence of perturbations that make it jump to another synchronous state. 
As pointed out by previous research on the Kuramoto model,~\cite{Dev12,Hin18} the expected time between two jumps is exponential in (i) the inverse of the natural frequencies' distribution width and (ii) the potential difference between the initial synchronous states and the closest $1$-saddle. 
It can also be related to (iii) the distance (in the state space) between the stable synchronous state and the closest $1$-saddle.~\cite{Tyl19} 
By the equivalence derived in Sec.~\ref{sec:equiv}, the expected time between jumps from a clustering pattern to another then follows the same exponential dependence (i)-(iii).

The equivalence is clearly seen in Fig.~\ref{fig:jumps}, where the angle trajectories in the two systems seem to be rescaled copies of each other, while they are obtained by two different simulations. 
The difference being that in the original Kuramoto model, the system jumps from the unique synchronous state to itself, by periodicity of the phase space, while in the Kuramoto model with $q^{\rm th}$-order coupling, the jumps occur between different clustered states. 

\subsection{Information storage}
It has been acknowledged~\cite{Kis05,Niy09} that the dynamics of some real-world systems can be described rather accurately by Eq.~\eqref{eq:kq}. 
In particular, it was proposed that such systems could be used as information storage,~\cite{Ash04,Ska11} where the value of each lower order parameter is a piece of information. 
Ref.~\onlinecite{Ska11} defines a forcing that allows to determine the clusterd state to which the system synchronizes and thus to control the value of $r_p$ for $1\leq p < q$. 
In this scope, the dynamical equivalence shown in Sec.~\ref{sec:equiv} indicates that the stability properties of the system does not depend on its states. 
This is of major importance for such an application as it guarantees that the reliability of the storage system does not depend on the information it contains, which would render such an application much more complicated.

\subsection{Generalization}
More generally, instead of non-oriented, homogeneous, all-to-all couplings, our argument can be straightforwardly extended to the Kuramoto model with interactions given by any graph, weighted or not, directed or not. 
Our argument also shows dynamical equivalence between the two following, more general versions of the Kuramoto model with higher-order coupling: 
\begin{align}
 \dot{\theta}_i &= \omega_i - \sum_{j=1}^n\sum_{\ell=1}^q\frac{K_\ell}{n}\sin\left[\ell p \cdot (\theta_i-\theta_j)\right]\, , \\
 \dot{\theta}_i &= p\omega_i - \sum_{j=1}^n\sum_{\ell=1}^q\frac{pK_\ell}{n}\sin\left[\ell \cdot (\theta_i-\theta_j)\right]\, ,
\end{align}
for $i\in\{1,...,n\}$ and $p\in\mathbb{Z}_{>0}$.

\section{Conclusion}\label{sec:conclusion}
The main consequence of the dynamical equivalence presented in this manuscript is that any property of the original Kuramoto model, Eq.~\eqref{eq:k1}, 
can be lifted to the Kuramoto model with simple $q^{\rm th}$-order coupling, Eq.~\eqref{eq:kq}. 
The only discrepancy being the multiplicity of the elements of the lifting from Eq.~\eqref{eq:kq} to Eq.~\eqref{eq:k1}.  

To summarize, we showed that the Kuramoto model with first-order coupling is dynamically equivalent to the Kuramoto model with simple $q^{\rm th}$-order coupling. 
As a matter of fact, any dynamical property of the latter can be derived from the corresponding property of the original Kuramoto model. 
The behavior of the Kuramoto model with higher-order coupling qualitatively changes only if at least two different coupling orders are considered. 
Clustering occurs in the Kuramoto model with simple $q^{\rm th}$-order couping because of the choice of coupling function. 
But the dynamics are blind to the clustering pattern as each synchronous states is dynamically equivalent. 

To take into account clustered states whose characteristics (linear stability, basin shape and size,...) differ, other models should be used. 
Some promising examples are, for instance, the more general Kuramoto model with $q^{\rm th}$-order coupling [Eq.~\eqref{eq:kuramotoQ}] or some dynamical systems with bounded confidence.~\cite{Lor07}

\section*{Acknowledgments}
This work has been supported by the Swiss National Science Foundation under grant 200020\_182050.


%

\end{document}